\documentclass[preprint,authoryear,12pt]{elsarticle}
\usepackage{graphicx}
\usepackage{amssymb} 
\usepackage{amsmath} 

\usepackage{color}
\newcommand{\ehg}[1]{#1}
\newcommand{\BE}{\begin{equation}}
\newcommand{\EE}{\end{equation}}
\newcommand{\BA}{\begin{eqnarray}}
\newcommand{\EA}{\end{eqnarray}}
\newcommand{\bx}{{\bf x}}
\newcommand{\by}{{\bf y}}
\newcommand{\bk}{{\bf k}}
\newcommand{\hG}{{\hat G}}
\newcommand{\hD}{{\hat D}}

\journal{}

\begin{document}

\begin{frontmatter}

\title{Spatial patterns of competing random walkers}

\author[1]{Emilio Hern\'{a}ndez-Garc\'{\i}a}
\ead{emilio@ifisc.uib-csic.es}
\author[2,3]{Els Heinsalu}
\author[1]{Crist\'{o}bal L\'{o}pez}
\address[1]{IFISC (CSIC-UIB) Instituto de F{\'\i}sica Interdisciplinar y Sistemas
Complejos, Campus Universitat de les Illes Balears, E-07122
Palma de Mallorca, Spain}
\address[2] {Niels Bohr International Academy, Niels Bohr Institute,
Blegdamsvej 17, DK-2100 Copenhagen, Denmark}
\address[3] {National Institute of Chemical Physics and
Biophysics, R\"avala 10, 15042 Tallinn, Estonia}

\begin{abstract}
We review recent results obtained from simple individual-based
models of biological competition in which birth and death rates
of an organism depend on the presence of other competing
organisms close to it. In addition the individuals perform
random walks of different types (Gaussian diffusion and
L\'{e}vy flights). We focus on how competition and random
motions affect each other, from which spatial instabilities and
extinctions arise. Under suitable conditions, competitive
interactions lead to clustering of individuals and periodic
pattern formation. Random motion has a homogenizing effect and
then delays this clustering instability. When individuals from
species differing in their random walk characteristics are
allowed to compete together, the ones with a tendency to form
narrower clusters get a competitive advantage over the others.
Mean-field deterministic equations are analyzed and compared
with the outcome of the individual-based simulations.
\end{abstract}

\begin{keyword}
Competition \sep Clustering \sep Pattern formation \sep
Individual-based model \sep Nonlocal interactions \sep Niche
space \sep Random walks \sep L\'{e}vy flights
\end{keyword}

\date{December 31, 2013}

\end{frontmatter}



\section{Introduction}

Competitive interactions are among the basic building blocks
shaping ecosystems and driving evolution. In its basic form,
competition refers to the situation in which two organisms
utilize the same resource to survive and grow. If one of them
consumes the resource, it is no longer available for the other,
which will then experience a decrease in growth capacity,
increase of mortality, or both. This basic effect of
competition as a limitation to growth is already present in
very early population dynamics mathematical models such as the
Verhulst or logistic equation \citep{Verhulst2006}, or in its
simplest generalization when taking into account spatial
dispersion, the Fisher or Fisher-Kolmogorov-Petrovskii-Piskunov
(FKPP) partial differential equation \citep{Okubo2001,
Murray2002,Mendez2014}. Resource limitation in such models
defines a carrying capacity which becomes the stable asymptotic
value of the population density. When space is taken into
account, this stable state advances into unpopulated regions as
a propagating front \citep{Fife1979,Okubo2001,
Murray2002,Mendez2014}.

It was recognized some time ago \citep{Britton1989,Sasaki1997}
that, in this situation in which the spatial dimension is
considered, the competition spatial range is an important
parameter than can change qualitatively population
dynamics.\ehg{ In particular, population density can arrange
into spatially periodic patterns when competitive interactions
occur within a finite region around the individuals
\citep{Britton1989,Sasaki1997}. This is in contrast with the
homogeneous configurations attained from the FKPP model that
assumes a strictly local competition.} In these works the
finite range of interaction enters the description via an
integral term, converting the FKPP model into an
integrodifferential equation, the nonlocal FKPP model. \ehg{In
this mathematical framework, the problem of individuals or
populations competing for resources in space becomes formally
identical to the one of different species
\citep{MacArthur1967}, or even different phenotypes of the same
species \citep[p. 534]{Roughgarden1979}, competing for
resources distributed in the so-called {\sl niche space}. This
is typically modeled by variations of the Lotka-Volterra
competition equations \citep{Volterra1926,Lotka1932}}. The
position of a species in niche space is defined as the set of
traits relevant to characterize resource utilization by this
species. Proximity in such niche space, usually assumed to be
onedimensional, implies utilization of similar resources which
implies stronger competition. \ehg{The question on whether
populations in space can remain homogeneously distributed in
the presence of competitive interactions or if rather they will
break in clusters with a typical spacing} is then related to
important issues in population and community ecology such as
the {\sl principle of competitive exclusion} or the question on
{\sl limiting similarity}
\citep{MacArthur1967,May1972,Abrams1983,Barabas2012,Leimar2013}.

The Lotka-Volterra equations or the related nonlocal FKPP model
have been intensively analyzed in recent years
\citep{Fuentes2003,Scheffer2006,Maruvka2006,Genieys2006,Pigolotti2007,
Leimar2008,Hernandez2009,Berestycki2009,Fort2009,Barabas2012,Leimar2013},
from which a reasonable understanding of the dynamics they
represent begins to emerge. At the same time, a different type
of approach has been developed in which the competing organisms
are modeled as individual agents following a stochastic
dynamics
\citep{HernandezGarcia2004,HernandezGarcia2005jphysc,Birch2006,Heinsalu2010,Heinsalu2012,Heinsalu2013}.
In this framework, the discrete nature of individual organisms
and fluctuations associated to the birth-death processes are
naturally taken into account. In addition, different types of
spatial motion of the organisms can be implemented as different
kinds of random walks. The dynamics of these {\sl
individual-based models} usually leads to results qualitatively
and in some cases even quantitatively similar to the
Lotka-Volterra approach. Some differences however arise related
to the fluctuating and discrete nature of the particle system
\citep{HernandezGarcia2004,Lopez2004,HernandezPA2005}.

In this Paper we review some of the basic results on the
dynamics of these stochastic and spatially extended discrete
models of competing individuals. More general approaches to
competition in spatial settings can be found for example in
\citet{Klausmeier2002} or \citet{Amarasekare2003}. Our focus
here is on the interplay between the competition interactions
and the random motion of the individuals. Two types of random
walks are considered: the standard Gaussian or Brownian random
walk, and a family of L\'{e}vy flights characterized by a
L\'{e}vy exponent $\mu$ which controls the probability of large
jumps \citep{Metzler2000,Klages2008,Mendez2014}. The L\'{e}vy
type of motion has been pointed out by its relevance in
efficient search strategies \citep{Benichou2011,James2011}, and
observed in a number of experimental studies
\citep{Dieterich2008,Matthaus2009,Matthaus2011,deJager2011}. We
show that, under suitable conditions, competitive interactions
lead to clustering and pattern formation. On the other hand,
one of the main effects of the random motions of the
individuals is to decrease and diffuse away inhomogeneities.
Then the occurrence or not of population clusters will depend
on the interplay between these two opposing forces. More
surprisingly, another effect of the type of motion is to
provide some competitive advantage to model organisms that are
identical in any other aspect: we will see that survival
competition is mediated by clustering, so that forming stronger
clusters, which depends on the random walk characteristics,
provides better chances for survival.

Reviewing results on the above two effects of mobility on
competition is the objective of this Paper. We try to put the
different results into a common framework. First we present
numerical simulations of the stochastic models, showing the
main phenomenology and highlighting the basic mechanisms by
means of heuristic arguments. Then we analyze suitable
Lotka-Volterra integrodifferential models able to capture in a
mean-field sense part of the observed phenomenology, pointing
out also the limitations of such approach. For definiteness we
concentrate in the situation of organisms competing for
resources in physical space. Most of the results have also an
interpretation in terms of species competition in niche space,
but this will only be briefly commented in the Conclusions
section.

In addition to summarizing previous results, and clarifying
previously proposed heuristic arguments, we add here some new
details on spatial structure of the competing populations and
its relationship with competitive advantage, together with a
study of extinction times of the different types of walkers.
The Paper is organized as follows: in Sect.~\ref{sec:model} we
describe the stochastic model of interacting particles and in
Sect.~\ref{sec:patterns} we present the results of the
numerical simulations, showing pattern formation, and their
interpretation both in heuristic terms and with the help of a
mean-field description. The limitations of this last approach
are also pointed out. In Sect.~\ref{sec:advantage} we put
together walkers with different types of motion, competing
between them, and focus on the phenomenon of competitive
advantage by pattern formation. Quantitative details on spatial
structure and extinction times are presented here, as well as a
theoretical analysis in terms of a two-species mean-field
description. Finally, in Sect.~\ref{sec:conclusions} we
summarize and discuss our conclusions.


\section{A model of competing bugs}
\label{sec:model}


We represent the state at time $t$ of a set of simple competing
organisms, reproducing asexually, by a set of $N(t)$ point-like
particles (they will also be referred as {\sl walkers}, {\sl
bugs}, or {\sl individuals}) at positions $\bx_i(t)$,
$i=1,2,...,N(t)$ in a twodimensional square domain of size
$L\times L$ (see \citet{Lopez2004} for simulations of a
onedimensional version). Periodic boundary conditions will be
assumed, and without loss of generality we take $L=1$.

The dynamics in the number of individuals includes birth and
death processes which are affected by the competitive
interactions. Namely, the bug labeled $i$ reproduces and dies
following Poisson processes of rates $r^i_b$ and $r^i_d$,
respectively, given by:
\BA
r^i_b &=& \mathrm{max} \left( 0, r_{b0} - \alpha N_R^i \right)
 \nonumber \\
r^i_d &=& r_{d0} + \beta N_R^i \ . \label{r-birthdeath}
\EA
$r_{b0}$ and $r_{d0}$ are the constant reproduction and death
rates experienced by an isolated bug. Competitive interactions
are introduced in the terms containing $\alpha$ and $\beta$:
The reproduction rate of individual $i$ decreases (we assume
$\alpha>0$) with the number of neighbors, $N_R^i$, that are at
a distance smaller than $R$ ($R < L$) from it. Analogously,
death rate increases ($\beta>0$) with the number of neighbors.
The newborns are placed at the same position as the parent bug,
introducing reproductive correlations in the system. The max
function in the first equation is needed to avoid rates (which
are probabilities per unit of time) to become negative.

\ehg{After each particle number update} with the birth-death
dynamics, the particles perform independent random walks. In
the case of Brownian walks all particles make jumps in random
directions and with sizes $l$ independently sampled from the
positive part of a Gaussian distribution of standard deviation
$\tilde l=(2\kappa \tau_0)^{1/2}$. This defines a diffusion
coefficient $\kappa$. It is convenient to perform the jumps
after each particle-number updating, and $\tau_0$ is the mean
time between such events. The case of L\'{e}vy motion
corresponds to the so-called L\'{e}vy flights
\citep{Klages2008,Metzler2000}: length $l$ of the jumps is
sampled from a L\'evy-type probability density that for large
$l$ decays as $\tilde l^{-1}(l/\tilde l)^{-\mu -1}$. $\mu
\in(0,2)$ is the anomalous exponent controlling the probability
of large jumps. The variance of the displacement is divergent,
but an anomalous diffusion coefficient $\kappa_\mu$ can be
defined from  $\tilde l$, the scale parameter: $\tilde
l=(2\kappa_\mu \tau_0)^{1/\mu}$. When particles move freely
without undergoing the birth-death process, their spatial
distribution develops fat tails at long distances, which is an
anomalous-diffusion behavior very different from the Gaussian
one \citep{Klages2008,Metzler2000}. The smaller $\mu$ the more
anomalous the diffusion process. When $\mu>2$, however, the
variance of the displacements becomes finite and the central
limit theorem guarantees that Gaussian statistics applies at
long times. We will see that the effect of the anomalous
diffusion is not so drastic in the presence of the birth-death
process or the finite interaction range $R$, since these
quantities provide effective time and space cut-offs before the
asymptotic behavior is attained. Both $\kappa$ and $\kappa_\mu$
will be denoted loosely as {\sl diffusivities}. The dynamics
defined in this way incorporates interactions in the
birth-death dynamics, whereas the random walks of the
individuals remain independent (except at the moment of birth
which fixes initial position). We are thus exploring mechanisms
which are different, and complementary, from the ones labeled
as {\sl active Brownian particles} \citep{Romanczuk2012}, in
which interactions affect motion and not the particle number.

The processes specified above are implemented here through the
Gillespie algorithm as described in \citet{Heinsalu2012}, where
additional details on the numerics can be found. Other
implementations are possible, which lead to similar results.
For example \citet{HernandezGarcia2004} used a discrete-time
approach nearly equivalent to the present one, except for very
large Poisson rates which become limited by the minimum time
imposed in the time discretization.

\ehg{Parameter $R$ is a central ingredient in our model, as it
imposes a finite range for the competitive interactions. In
fact we will see that the periodicity of the spatial patterns
that are described below is directly related to $R$. In this
paper we will focus in the situation in which $R$ is much
smaller than system size $L$ (typically we take $R=0.1$ as
compared to $L=1$) so that boundary conditions become
relatively irrelevant. On the other hand, periodic-pattern
formation requires $R$ to be substantially larger than the
sizes of particle clusters (see Sect. \ref{sec:observations1}).
This will be achieved by considering small values of the
diffusivities introduced above.}


\section{Clustering of particles and pattern formation}
\label{sec:patterns}


\subsection{The competitive clustering instability}
\label{sec:observations1}

In simulations of Brownian or L\'{e}vy bugs with a sufficiently
large $\kappa$ or $\kappa_\mu$, respectively, a nearly
homogeneous distribution of particles is attained which remains
fluctuating but statistically steady. The expected density in
such situation can be estimated from the requirement of a
balance between births and deaths. From Eq.
(\ref{r-birthdeath}) and neglecting the max condition (which is
correct if we neglect fluctuations) this is achieved when
$N_R^i\approx (r_{b0}-r_{d0})/(\alpha+\beta)$. In a homogeneous
situation the expected density $\rho_0$ is the same everywhere,
and since $N_R^i$ counts the number of particles in a circle of
area $\pi R^2$, we have
\BE
\rho_0=\frac{\Delta}{\gamma \pi R^2}
\label{homo}
\EE
where the notation
\BE
\Delta\equiv r_{b0}-r_{d0} \ \ \ \textrm{and} \  \
\gamma\equiv\alpha+\beta \label{Deltagamma}
\EE
has been introduced. This estimation is correct for
sufficiently large $\Delta$. For small $\Delta$ the expected
density is small and stochastic extinctions are likely to occur
from fluctuations in the particle number. In this case, in
average, the expected density becomes smaller than (\ref{homo})
and even vanishes well above $\Delta=0$
\citep{HernandezGarcia2004,Lopez2004,HernandezPA2005}.

\begin{figure}[h] \centering
\includegraphics[width=0.32\textwidth]{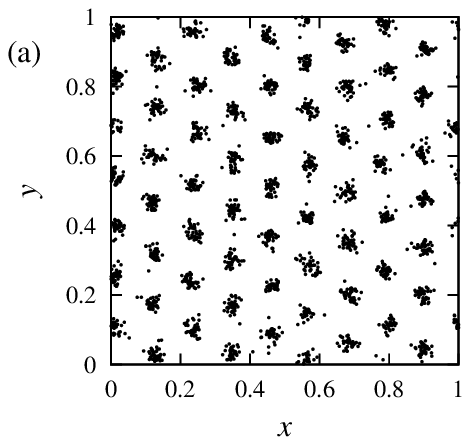}
\includegraphics[width=0.32\textwidth]{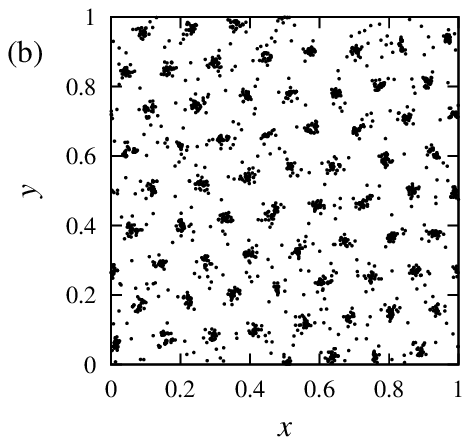}
\includegraphics[width=0.32\textwidth]{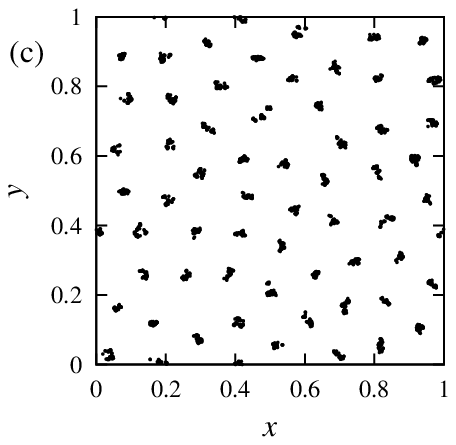}
\caption{Snapshots of configurations of competing bugs in statistically steady states. In all panels
$r_{b0}=1$, $r_{d0}=0.1$, and $R=0.1$. In (a) and (b) competition enters only in the birth rate, with
$\alpha=0.02$ and $\beta=0$. (c) considers competition only in the death rate, $\alpha=0$, $\beta=0.02$.
Motion in panels (a) and (c) is of the Gaussian type, with $\kappa=10^{-5}$. L\'{e}vy
flights with $\mu=1$ and $\kappa_\mu=56\times 10^{-5}$ occur in (b).}
\label{Fig-configs}
\end{figure}

But the most interesting dynamics occur when diffusivities are
sufficiently small. In this situation the homogeneous density
breaks down and an approximately periodic arrangement of
particle clusters arise, both for Brownian as for L\'{e}vy
walkers (see Fig. \ref{Fig-configs}). The symmetry of the
resulting pattern is roughly hexagonal. Configurations such as
the ones in Fig. \ref{Fig-configs} can be analyzed with
different types of correlations and structure functions
\citep{HernandezGarcia2004,Lopez2004,Birch2006,Ramos2008,Heinsalu2010}.
We delay this type of analysis until section
\ref{sec:advantage}, where it will be used to characterize
competition between different types of walkers. In the
following we just comment on the main aspects: Most of the
parameters used in the three simulations illustrated in Fig.
\ref{Fig-configs} are equal, the only differences being the use
of Brownian (panels (a) and (c)) or L\'{e}vy (panel (b)) jumps,
and the presence of competition only in the birth rate (panels
(a) and (b)) or only in the death rate (panel (c)). In the
three cases the number of particles is similar (fluctuating
close to 2600) as well as the number of well-formed clusters
(63, 68 and 62 for panels (a), (b) and (c), respectively). The
main difference is in the shape of the clusters and in their
ordering: In the L\'{e}vy case, because of the larger
probability of long jumps, there are much more {\sl solitary}
particles \citep{Heinsalu2010,Heinsalu2012} that jump into the
regions between clusters, remaining there for a short time
before dying or jumping again. In the Gaussian case clusters
are much more compact. When competition affects only the death
rate (panel (c)) clusters are narrower and are arranged in a
more disordered way. The reasons for this will be commented at
the end of subsection \ref{sec:limitations1}.

\begin{figure}[h] \centering
\includegraphics[width=.5\textwidth]{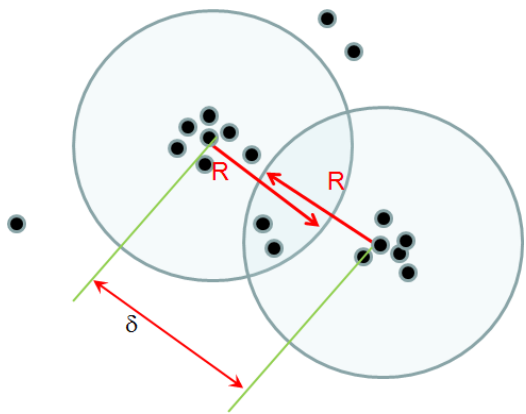}
\caption{A sketch illustrating the mechanism of the clustering instability:
two main clusters of particles are displayed, separated by a distance $\delta$ between $R$ and $2R$.
The circles approximate the respective competition ranges. Particles in the intersection between the
two ranges, the {\sl death zone}, experience competition from bugs in the two clusters.}
\label{Fig-qualit}
\end{figure}

One can understand the instability of the homogeneous
distribution by some qualitative arguments that clearly
indicate that it is a consequence of the competitive
interactions: Let us consider the bugs initially distributed
homogeneously in space with the density $\rho_0$, so that
deaths and births are balanced. When there are fluctuations or
perturbations that enhance the density at points separated by a
distance larger than $R$ but smaller than $2R$, death will
become more probable than reproduction in the area in between
these density maxima. This is so (see Fig. \ref{Fig-qualit})
because of Eqs.~(\ref{r-birthdeath}) and of the fact that in
this zone an individual experiences the competition with the
organisms from at least two of the density maxima, whereas in
each density maximum the competition takes place only between
the individuals in the same maximum (other maxima are simply
out of the interaction range $R$). As a consequence, the
density will decrease in between the density maxima. This in
turn releases competitive pressure on the maxima, which will
tend to grow, and then start to form periodically located
clusters (at a distance $fR$, with $1<f<2$) and close a
positive feedback loop that will finally eliminate all
organisms in these {\sl death zones} between the clusters. The
only mechanism that can stop this process is particle
diffusion, if it occurs fast enough to redistribute the bugs
before the instability concentrates them.

The above qualitative argument clearly identifies the
competitive interactions as responsible for the pattern forming
instability. However it requires, besides small diffusivities
to guarantee that clusters do not extend into the {\sl death
zones}, that there is a sufficiently sharp distinction between
the interior and the exterior of the competition area. The
precise meaning of these requirements will be clarified in the
next subsection. We close this one by noticing that the number
of particles in the system in which clustering occurs is higher
than what would be allowed by the homogeneous distribution. To
see this we recall the previous calculation of $\rho_0$ in
which $N_R^i$ bugs are distributed in an area $\pi R^2$. Now
(Fig. \ref{Fig-configs}) the same number of bugs is
concentrated in small clusters in the center of regions which
are approximately hexagons of apothem $fR/2$. Since the area of
such hexagons is much smaller than the area of the circle of
radius $R$ (this is so as far as $f<2.199$) the mean density is
higher and we have more particles in the system. The hexagonal
shape is not necessary for the argument; it is enough that
clusters are associated to regions of width $fR$, with $f<2$,
instead of circles of diameter $2R$. By clustering together and
leaving in between sufficient space free from competitors (i.e.
leaving empty the death zones), the bugs experience a smaller
effective competition and can survive in larger numbers.

\subsection{Mean-field approach}
\label{sec:theory1}

Let us analyze a deterministic description of the system, which
is enough to formalize the heuristic arguments given above.
Denoting the mean local density of walkers by $\rho(\bx,t)$,
different mathematical arguments
\citep{HernandezGarcia2004,Birch2006} can be put forward to
justify that, if statistical fluctuations are neglected, the
following dynamics approximates the mean evolution of the
density:
\BE
\frac{\mathrm{\partial} \rho(\bx,t)}{\mathrm{\partial} t} =
M(\bx,t) \rho(\bx,t)  + D \rho(\bx,t) \, .
\label{1eq}
\EE
Here, $D$ stands for the diffusion-like operator representing
the effect of mobility on the particle density. For Brownian
walkers, it is the simple diffusion operator
$D=\kappa\nabla^2$, whereas in the L\'{e}vy case $D=\kappa_\mu
\nabla^\mu$ involves the fractional derivative of order $\mu$
\citep{Metzler2000,Klages2008,Mendez2014}. The quantity
$M(\bx,t)$ is
\BE
M(\bx,t) \equiv \Delta - G_\bx* \rho\ ,
\EE
which simply the difference between birth and death rates from
Eq. (\ref{r-birthdeath}). $\Delta$, defined in
(\ref{Deltagamma}), is a net linear growth rate. The symbol
$G_\bx*$ denotes the convolution product with a kernel
$G(\bx)$, i.e.,
\BE
G_\bx* f \equiv \int \mathrm{d} \by \, G(\bx-\by) f(\by) \, ,
\EE
where the integration is over all the system domain. In our
case, in which interactions enter the dynamics via
Eq.~(\ref{r-birthdeath}) which counts the number of organisms
in a two-dimensional neighborhood of radius $R$, the kernel
$G(\bx)$ is:
\BE
\begin{aligned}
G(\bx) =
\begin{cases}
\gamma \, , ~ \mathrm{if} ~ |\bx|<R \, ,  \\
0 \, , ~ \mathrm{elsewhere} \, ,
\end{cases}
\label{kernel}
\end{aligned}
\EE
being $\gamma$ the competition slope defined in
(\ref{Deltagamma}). The above notation is general enough to be
useful in cases of arbitrary dimensionality and when the
interaction differs from (\ref{kernel}), for example if it is
weighted with the distance
\citep{Birch2006,Pigolotti2007,Hernandez2009,Pigolotti2010}.

The type of description given by Eq.~(\ref{1eq}) turns out to
be accurate where the densities are large enough to neglect
fluctuations. We also mention that the use of the operators
$\nabla^2$ and $\nabla^\mu$ defining $D$ is accurate only at
sufficiently large time and spatial scales. Numerical
simulation of Eq. (\ref{1eq}) for low diffusivities correctly
exhibits the pattern forming instability leading to an
hexagonal pattern with a periodicity similar to the one
observed in the particle simulations
\citep{HernandezGarcia2004,Lopez2004,Lopez2007a}.

We now use this continuous description to understand
analytically the pattern forming instability. To do so, we
introduce some notation: we call $\hG_\bk$ the Fourier
transform of $G(\bx)$, i.e., $\hG_\bk = \int \mathrm{d} \bx \,
e^{i \bk \cdot \bx} G(\bx)$, and $\hG_0 \equiv \hG_{\bk = {\bf
0}} = \int \mathrm{d} \bx \, G(\bx)$. For our two-dimensional
kernel (\ref{kernel}) these functions become:
\BA
\hG_\bk &=& 2 \gamma \pi R^2 \frac{J_1 (kR)}{kR} \nonumber \\
\hG_0   &=& \gamma \pi R^2 \ , \label{GkG0}
\EA
with $J_1$ being the first-order Bessel function and $k =
|\bk|$. Also, the action of the operator $D$ in the Fourier
representation involves simple multiplications: $\hD(k)=
-\kappa k^2$ in the Brownian case or $\hD(k)= -\kappa_\mu
k^\mu$ under L\'{e}vy motions.

The spatially homogeneous steady solution of Eq.~(\ref{1eq}) is
precisely $\rho_0$ in (\ref{homo}), which is independent of the
particular form of the operator $D$. Now we perturb $\rho_0$
with harmonic functions and look at the growth rates of such
perturbations to assess stability:
\BE
\rho(\bx,t) = \rho_0 + \epsilon \, e^{\lambda(\bk) t}
e^{i\bk\cdot\bx}
 \, . \label{perturbation1}
\EE
Linearizing in the small perturbations $\epsilon$ one finds
\BE
\lambda(\bk)=\hD(k)-\Delta\frac{\hG_\bk}{\hG_0} \ .
\label{lambdak}
\EE
Since $\hat D(k)<0$, we see that $\lambda(\bk)$ will remain
negative, and then $\rho_0$ will remain stable, when $\hG_\bk$
is positive $\forall~\bk$. When $\hG_\bk$ is negative for some
$\bk$, however, the second term can destabilize the homogeneous
solution if the mobility term characterized by $\hD(k)$ is not
sufficiently negative at these wavenumbers. The change from
negative to positive sign in $\lambda(\bk)$ will occur first
for the wavenumber $k_c=|\bk_c|$ for which $\lambda(\bk)$ is
maximum. The instability will develop in a periodic pattern
which, at least close enough to the instability
\citep{Cross1993}, will have a periodicity $\delta$ determined
by the growing wavenumber, $\delta=2\pi/k_c$. Equation
(\ref{lambdak}) clearly expresses the first effect of particle
mobility characterized by $\hD(k)$ on the competitive dynamics:
$\hD(k_c)<0$ will stabilize the instability produced by the
negative values of $\Delta \hG_\bk /\hG_0$ at $|\bk|\approx
k_c$ until they are too large. From the expressions
$\hD(k_c)=-\kappa k_c^2$ or $-\kappa_\mu k_c^\mu$ we can
identify these quantities as the rates of particle-escape out
of the periodic structures of wavenumber $k_c$: increased
diffusivities enhance the flux of particles from the clusters
towards the death zones, thus working against the pattern
formation process (and reducing the mean density).

The qualitative requirement discussed in subsection
\ref{sec:observations1} of {\sl a sufficiently sharp
distinction between the interior and the exterior of the
competition area} becomes now formalized: the sharp distinction
occurs when the Fourier transform $\hG_\bk$ of the kernel takes
negative values. For example, for kernels of the form $G(\bx)
\propto \exp(-|\bx/\sigma|^p)$, the Fourier transform has
negative components \citep{Bochner1937}, and then they are
sufficiently sharp to produce the competitive instability, in
the platykurtic situation $p>2$ (note that the flat-top kernel
in (\ref{kernel}) corresponds to $p\rightarrow\infty$). Under
kernels with $p\le 2$ (which include the Gaussian and the
exponential kernels) $\hG_\bk$ is positive and then the
homogeneous distribution will remain stable.

\begin{figure}[h] \centering
\includegraphics[width=\textwidth]{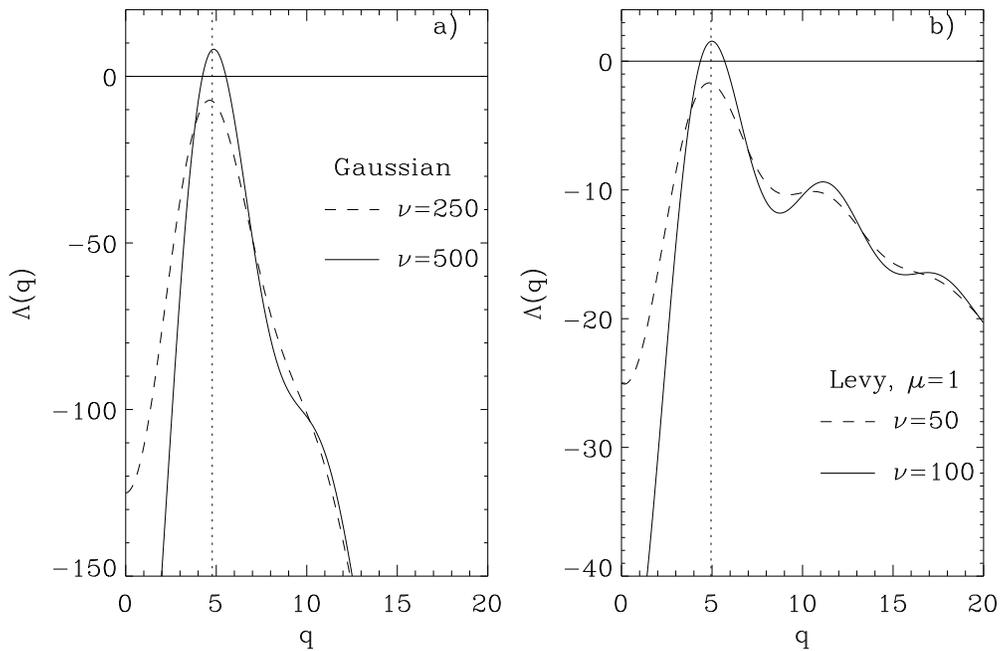}
\caption{The linear growth rate from Eq. (\ref{growthrate}). Panel a) displays the Gaussian case (formally
$\mu=2$ in Eq. (\ref{growthrate})) for two values of $\nu$, one below and another above
the cluster instability occurring at $\nu_c=370.384$ (see Table \ref{table1}). Panel b) is for
the L\'{e}vy case with $\mu=1$ for which $\nu_c=76.1997$. In both panels the vertical dotted line indicates the
critical wavenumber ($q_c$ from Table \ref{table1}) at which instability first occurs.}
\label{Fig-growth}
\end{figure}

We now return to the specific competition kernel given by
(\ref{kernel}) and (\ref{GkG0}). We introduce dimensionless
versions of the wavenumber, $q\equiv kR$, and of the growth
rate, $\Lambda(q)\equiv R^\mu\lambda(\bk)/\kappa_\mu$ for the
L\'{e}vy case. The corresponding expressions in the Gaussian
case are obtained by using $\mu=2$ and replacing $\kappa_2$ by
$\kappa$. Equation (\ref{lambdak}) then reads
\BE
\Lambda(q)= -q^\mu-\nu\frac{J_1(q)}{q} \ ,
\label{growthrate}
\EE
where the dimensionless combination $\nu\equiv
2R^\mu\Delta/\kappa_\mu$ has been introduced. We plot Eq.
(\ref{growthrate}) in Fig. \ref{Fig-growth} for the Gaussian
case (formally obtained by substituting $\mu=2$) and for a
L\'{e}vy case with $\mu=1$. \ehg{We have that pattern formation
occurs when $\nu>\nu_c$, which can be achieved by increasing
$R$ or $\Delta$, or reducing the diffusivity $\kappa_\mu$.}

The critical wavenumber $k_c=q_c/R$ for which instability first
occurs can be found by solving simultaneously the equations
$\Lambda(q_c)=0$ and $\Lambda'(q_c)=0$, which implement the
conditions of change of sign in the growth rate and of maximum
of the growth rate, respectively. Using these conditions and
properties of derivatives of Bessel functions one obtains the
critical value of $\nu$: $\nu_c=-q_c^{\mu+1}/J_1(q_c)$ and the
equation for the critical wavenumber:
\BE
q_c\frac{J_2(q_c)}{J_1(q_c)}=-\mu \ .
\EE
Numerical solution of this equation gives $q_c=q_c(\mu)$ ($=
k_cR$) from which the expected periodicity of the pattern is
$\delta=2\pi R/q_c$. Table \ref{table1} presents these
quantities for several values of $\mu$ and for the Gaussian
case. In agreement with the previous heuristic discussion, we
see that pattern periodicity (last column in the Table) is
always of the form $\delta=fR$, with $1<f<2$.

\subsection{Limitations of the deterministic description}
\label{sec:limitations1}

The predicted periodicity is in good agreement with numerical
simulations \citep{HernandezGarcia2004,Heinsalu2010}, which
confirms the relevance of the nonlocal KPPP equation
(\ref{1eq}) to describe the particle system. There is only a
weak dependence of the wavenumber on diffusivities which is not
predicted by the mean-field approach. But there are other
features that are not captured by this deterministic
description:

First, as already commented, the homogeneous solution $\rho_0$
in (\ref{homo}) overestimates the mean density when $\Delta$ is
small. Because of number fluctuations, the particle system
becomes extinct when $\Delta$ is still positive. This
transition is of the Directed Percolation type
\citep{Lopez2007b}, qualitatively different from the mean-field
transition to the extinct state predicted by Eq. (\ref{1eq}) at
$\Delta=0$, but that can be recovered by incorporating
multiplicative noise terms to it \citep{Ramos2008}.

Second, even if pattern periodicity is correctly described by
(\ref{1eq}), cluster width is not. Extreme discrepancy occurs
when $R$ grows to be of the order of the system size, in which
case solutions of (\ref{1eq}) give a patch occupying the whole
system, whereas the particles in the discrete model remain
aggregated in a single small cluster
\citep{HernandezGarcia2005jphysc,Heinsalu2012}. In fact cluster
width, as in \citet{Young2001}, is determined by reproductive
correlations: it is essentially
\citep{HernandezPA2005,Heinsalu2012} the distance
$(2\kappa\tau_d)^{1/2}$ or $(2\kappa_\mu\tau_d)^{1/\mu}$
traveled during a particle lifetime $\tau_d\approx
(r_d^i)^{-1}$ (see \citet{HernandezGarcia2005jphysc} for a
refinement that takes into account the lifetime of the {\sl
lineage} of a particle). \ehg{\citet{Olla2012} discuses how
different mobility types affect reproductive correlations in
noninteracting particle systems.}

As a third deficiency of the deterministic modeling we point
out that, although for the parameter values used in panels (a)
and (c) of Fig. \ref{Fig-configs} the corresponding instances
of Eq. (\ref{1eq}) are identical, particle configurations are
indeed quite different. As noted before, the spacing between
clusters is very similar, but clusters in (c) are narrower and
are arranged in a more irregular way. Both differences have the
same origin: In Eq. (\ref{1eq}) only the difference between the
birth and death rates enter. But for the stochastic model the
absolute value of the rates is also relevant. For example, if
we balance the birth and death rates (neglecting the
diffusivities) to obtain steady values $r_b^{st}$ and
$r_d^{st}$, we find
$r_b^{st}=r_d^{st}=r_{d0}+\beta\Delta/(\alpha+\beta)$, which is
an increasing function of $\beta$. Thus, giving increasing
weight to competition in the death term while keeping other
parameters such as $\Delta$ or $\gamma$ unchanged increases the
value of the rates in steady state. The smaller width of the
clusters in panel (c) of Fig. \ref{Fig-configs} with respect to
the ones in panel (a) is a consequence of the smaller lifetime
of the bugs because of the increased steady death rate. Also,
the rate parameter in a Poisson process not only fixes the mean
number of events per unit of time, but also its variance. Then
higher rates produce more fluctuating processes
\citep{Heinsalu2012}, even if the mean value of the difference
between birth and death rates remains constant. This explains
the larger irregularity of the pattern in panel (c) of Fig.
\ref{Fig-configs} with respect to panel (a) which has $\beta=0$
and then smaller steady value of the rates. This effect, as
others linked to fluctuations, is absent from Eq. (\ref{1eq}).
\ehg{Another possible consequence of fluctuations that we have
not studied in detail could be the appearance of quasipatterns
excited by noise \citep{Butler2011} even in parameter regions
in which the deterministic description predicts a homogeneous
state.}

\section{Clustering as a competitive advantage}
\label{sec:advantage}

\subsection{Two-species competition}
\label{sec:observations2}

In Sect. \ref{sec:patterns} we have seen how competitive
interactions with some features lead to particle clustering and
pattern formation, and how particle mobility works against this
and shapes the pattern characteristics. In this section we
present a different phenomenon: competition advantage provided
by clustering. We consider competition between two types of
bugs (we can think of them as pertaining to two species),
identical in every aspect except in the type of motion. We will
see that despite having the same birth and death rates, and
identical direct competition properties, the different
clustering characteristics associated to the different type of
random walk provide a competitive advantage to one of the
species (roughly the one forming the narrowest clusters). When
competing together one of the species survives and the other
gets extinct.

The model set-up is as before, except that initially half of
the individuals are Brownian random walkers characterized by
the diffusion coefficient $\kappa$ and the other half are
L\'evy random walkers characterized by the exponent $\mu$ and
the anomalous diffusion coefficient $\kappa_\mu$. Birth and
death rates are the same as before for the two species, Eq.
(\ref{r-birthdeath}), with $N_R^i$ counting all neighbors
(i.e., of any type) of bug $i$. When reproducing, offspring is
of the same type as the parent walker.

When $\kappa$ and $\kappa_\mu$ are sufficiently large
\citep{Heinsalu2013}, walkers appear to be distributed in an
unstructured way, as when a single species is present. Local
fluctuations occur around an homogeneous mean without a
long-lasting pattern forming. The population numbers of the two
types of walkers widely fluctuate in antiphase, whereas the
total number weakly fluctuates close to a well-defined value.
In this situation in which both species are highly diffusive,
the two types of walkers become well mixed, so that there is no
difference in the neighborhood seen by the individuals of one
or the other type. Thus, effectively, the two species become
equivalent. The observed neutral fluctuations that conserve the
total number were expected from the stochasticity of the
reproduction-death process and the equivalence between the
species. After some time, however, \ehg{neutral fluctuations
can create diffuse patches dominated by a single species, and
eventually a particular neutral fluctuation could eliminate all
individuals of a given type}, after which only the other
species will remain in the system. The winner is not a better
competitor, it is just determined by chance.

The situation becomes different when $\kappa$ or $\kappa_\mu$
decrease \citep{Heinsalu2013}. When one of these quantities
becomes sufficiently small the corresponding species begins to
form clusters, in a manner similar to the one discussed in
Sect. \ref{sec:patterns}. At this moment the population of the
other species begins to decrease until complete extinction. The
possibility to develop clusters, determined by the type of
motion, has given a competitive advantage to one of the
species. When the simulation is started with the two types of
bugs randomly mixed and the values of $\kappa$ and $\kappa_\mu$
are such that both species form periodic patterns when alone,
then clusters begin to form and the bugs compete, until
typically only one of the species survives. The observed
outcome at some particular values of the birth/death parameters
is summarized in Fig.~\ref{Fig-Phase}.  \ehg{Variables in the
axes are $|\hat D_\mu(k_L)|=\kappa_\mu k_L^\mu$ (with $\mu=1$)
and $|\hat D_2(k_B)|=\kappa k_B^2$. $k_B$ is the critical
wavenumber $k_c$ for which instability first occur in the
Brownian walker case, whereas $k_L$ is the critical $k_c$ in
the L\'{e}vy case. Both quantities can be read-off from Table
\ref{table1}. This scaling of the axes is suggested by the
analytical approach of subsection \ref{sec:theory2}. For now it
is enough to keep in mind that these quantities are
proportional to the diffusivities $\kappa_\mu$ (with $\mu=1$)
and $\kappa$.} We see that when one of the diffusivities is
sufficiently small, the corresponding species wins. There is
also a small region in which coexistence persists during all
the long simulation times. At variance with the case of large
diffusivities, here the coexisting species are not mixed but
occupy segregated clusters, and both particle numbers, not just
their sum, fluctuate around well-defined means. There are also
some transition regions where the outcome depends on the
initial condition or the particular random realization. When
approaching these transitions zones, the extinction times of
the loosing species grow and finally diverge, signaling a
change of winner. This will be quantified in subsection
\ref{sec:times}.

By comparing visually the type of patterns formed by each of
the two species when they are alone in the system one
recognizes that, roughly, the organisms forming more
concentrated and populated clusters (which is associated to
lower diffusivities) are more successful when put into
competition with the other species. We quantify this
observation in  subsection \ref{sec:radial} by comparing
spatial patterns in terms of radial distribution functions.
Before that we mention that the general trends of winner and
loosing species can be understood from the heuristic arguments
set up in Sect. \ref{sec:observations1}: the existence of {\sl
death zones} in between particle groups would develop the same
clustering instability as before for both species, but this is
opposed by the flux of particles out of the clusters because of
their random walks. The species for which this flux is higher,
which manifest in wider clusters or in its absence, will visit
more frequently the {\sl death zones} and this gives them a
clear competitive disadvantage with respect to the ones that
remain in well concentrated clusters. The next subsection will
quantify this relationship between cluster compactness and
competition outcome and, after subsection \ref{sec:times}
devoted to extinction times, we will formalize these intuitive
ideas in \ref{sec:theory2} to provide some quantitative
criterion about competition outcome.


\subsection{Spatial structures in single-species simulations}
\label{sec:radial}


To understand the conditions favoring the surviving of one or
the other species, we return in this subsection to the
situation in which only one species is present in the system
and investigate their radial distribution function $g(r)$. It
is defined by using the formula $dn=(N/L^2)g(r)2\pi r dr$ which
involves the number of particles $dn$ at a distance between $r$
and $r+dr$ from a specific one, and then averaging over all
these specific particles and over long times. $N$ is the total
number of particles in the simulation domain of area $L^2$.
$g(r)$ describes \citep{Dieckmann2001} how the density varies
with the distance from a given particle with respect to the
uniform distribution (characterized by $g(r)=1$).

The value of $g(r \to 0)$ gives an idea of the number of close
neighbors surrounding any individual, i.e., it indicates the
increase in local density, induced by the presence of one of
the organisms, with respect to the uniform density. The
presence of the first peak of $g(r)$ at $r \neq 0$ indicates
the formation of a periodic pattern with periodicity $r$. The
value of $g(r)$ in between the first two peaks measures the
density between clusters compared to the uniform distribution.
The higher and narrower the peak at $r \to 0$, the more
concentrated are the organisms in the clusters.

\begin{figure}[t]
\centering
\includegraphics[width=\textwidth]{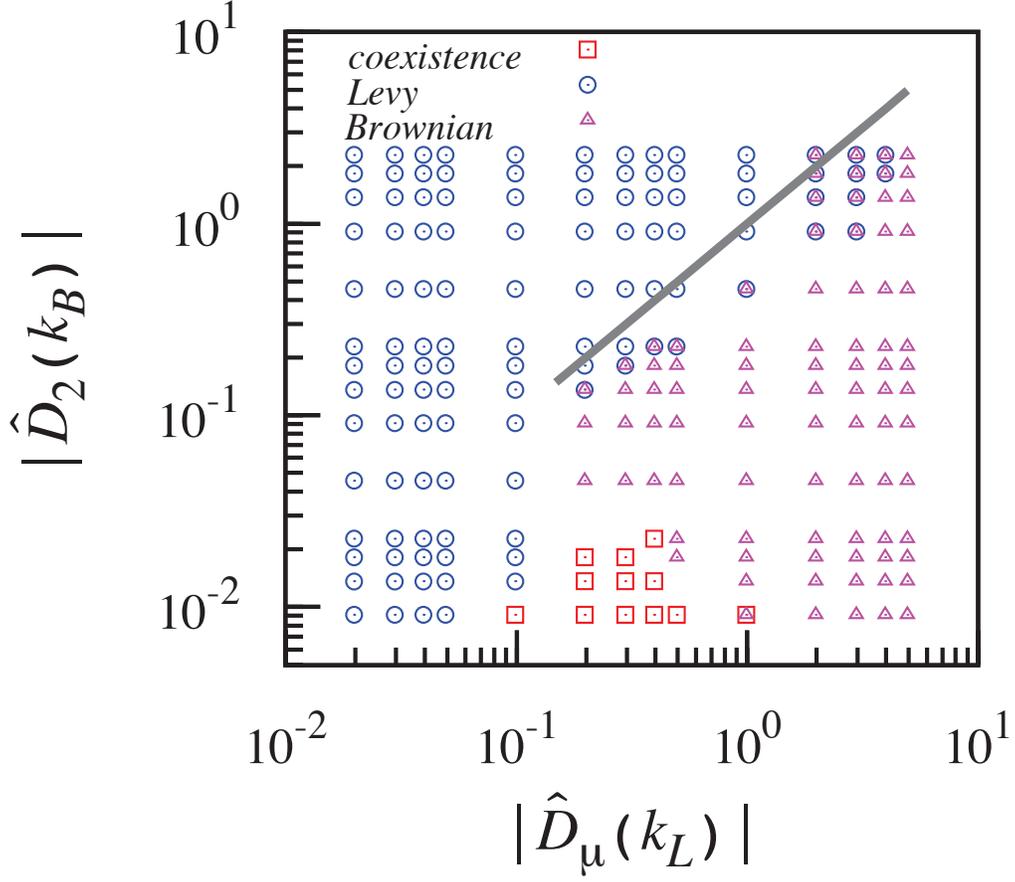}
\caption{Outcome of competition between Brownian and L\'evy walkers (with $\mu=1$) at $r_{b0}=1$,
$r_{d0}=0.1$, $R=0.1$, $\alpha=0.02$ and $\beta=0$. Symbols indicating which is the winner in the
competition are displayed as a function of the quantities
$|\hD_\mu(k_L)|=\kappa_\mu (q_c/R)^\mu=49.47~ \kappa_1$ for the
L\'{e}vy walkers, and  $|\hD_2(k_B)|=\kappa~ (q_c/R)^2=2284 \kappa$ for the Brownian ones
(critical wavenumber values $q_c$ are from Table \ref{table1}). Depending on the diffusivity
values $\kappa$ and $\kappa_1$ either Brownian
or L\'evy walkers win, or coexistence occurs.
Each point reflects the outcome of 25 realizations. Superimposed symbols
indicate that individual realizations end in different states. The main diagonal
sketched by the straight line is the prediction for the change in winner obtained in subsection
\ref{sec:theory2}.
}
\label{Fig-Phase}
\end{figure}

\begin{figure}[t]
\centering
\includegraphics[width=\textwidth]{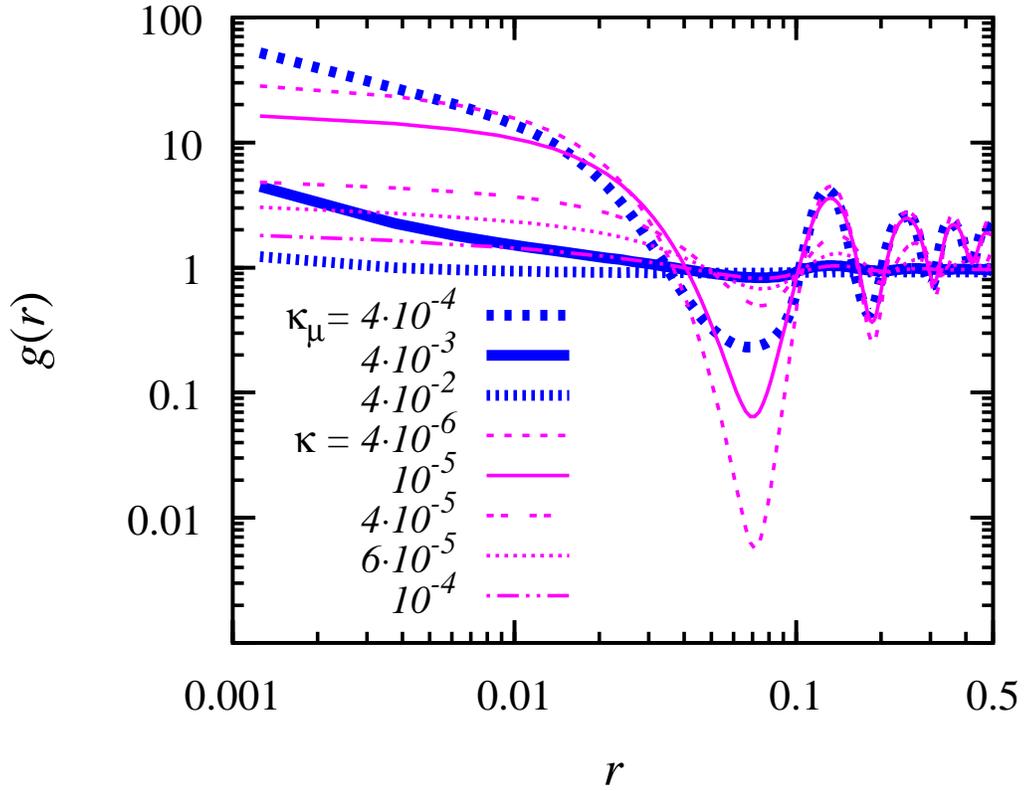}
\caption{Radial distribution function for single-species systems consisting
of L\'evy walkers characterized by different values of the anomalous diffusion
coefficient $\kappa_\mu$, with $\mu=1$, or of Brownian walkers characterized by different
values of diffusion coefficient $\kappa$. Other parameters as in Fig. \ref{Fig-Phase}.
}
\label{Fig-Radial}
\end{figure}

In Fig.~\ref{Fig-Radial} we depict $g(r)$ for the two types of
walkers and for different values of $\kappa$ and $\kappa_\mu$.
The three chosen values of $\kappa_\mu$ are intended to
represent the three qualitatively different situations seen in
Fig.~\ref{Fig-Phase} for the case of competing walkers: for
$\kappa_\mu = 4 \times 10^{-4}$, which corresponds in the
figure to $|\hD_\mu(k_L)|=0.0198$, L\'evy walkers always win;
for $\kappa_\mu = 4 \times 10^{-2}$ ($|\hD_\mu(k_L)|=1.98$)
Brownian walkers win; for $\kappa_\mu= 4 \times 10^{-3}$
($|\hD_\mu(k_L)|=0.198$), depending on the value of $\kappa$,
there is either coexistence of the two species or L\'evy or
Brownian walkers win the competition.

The value of $g(r \to 0)$ for the L\'evy walkers with the
smallest value of $\kappa_\mu$ or $|\hD_\mu(k_L)|$ plotted in
Fig. \ref{Fig-Radial} is higher than for Brownian walkers with
any of the values of $\kappa$ considered. This means that
L\'{e}vy clusters are relatively more populated than Brownian
ones in this range of parameters (see also
\citet{Heinsalu2010,Heinsalu2012}). Instead, for the largest
value of $\kappa_\mu$ or $|\hD_\mu(k_L)|$ there is no
clustering of L\'evy walkers ($g(r) \approx 1$) and the value
of $g(r \to 0)$ is always larger for the systems of Brownian
walkers with the studied values of $\kappa$. In the first case
the L\'evy walkers win, whereas in the second case the Brownian
walkers win, suggesting that in the two-species model the
survival is favored by the stronger clustering.

The situation is a bit more complex for the intermediate value
$\kappa_\mu = 4 \times 10^{-3}$ ($|\hD_\mu(k_L)|=0.198$). As
can be seen from the radial distribution function
(Fig.~\ref{Fig-Radial}), L\'evy walkers do not form a clear
periodic structure, although there is still weak local
clustering as indicated by $g(r \to 0) > 1$. But clustering is
even weaker for the Brownian walkers with $\kappa = 10^{-4}$
($|\hD_2(k_B)|=0.23$). Because the value of $g(r \to 0)$ is
sufficiently higher in the system of L\'evy walkers, the latter
ones win the competition. When reducing the Gaussian
diffusivity to $\kappa = 6 \times 10^{-5}$
($|\hD_2(k_B)|=0.137$) the difference of $g(r \to 0)$ between
the L\'evy and Brownian systems is rather small and the winner
of the competition is a random event. Reducing further $\kappa$
to $\kappa = 4 \times 10^{-5}$ ($|\hD_2(k_B)|=0.091$) the
values of $g(r \to 0)$ are approximately equal in the L\'evy
and Brownian systems; in this case, however, a clearer cluster
periodicity in the Brownian case (larger value of the the first
peak of $g(r)$ at $r\neq 0$) seems to give them advantage over
the L\'{e}vy ones.

Following this observation that competition success is attained
by the motion leading to the strongest and clearest clustering,
one would guess that reducing further $\kappa$ to $\kappa = 4
\times 10^{-6}$ ($|\hD_2(k_B)|=0.0091$) with the L\'{e}vy
diffusivity still at $\kappa_\mu = 4 \times 10^{-3}$ the
Brownian walkers would win. However, as indicated by
Fig.~\ref{Fig-Phase}, this is not the case. Instead,
coexistence occurs. This situation will be commented at the end
of next subsection.


\subsection{Extinction times}
\label{sec:times}

\begin{figure}[t]
\centering
\includegraphics[width=\textwidth]{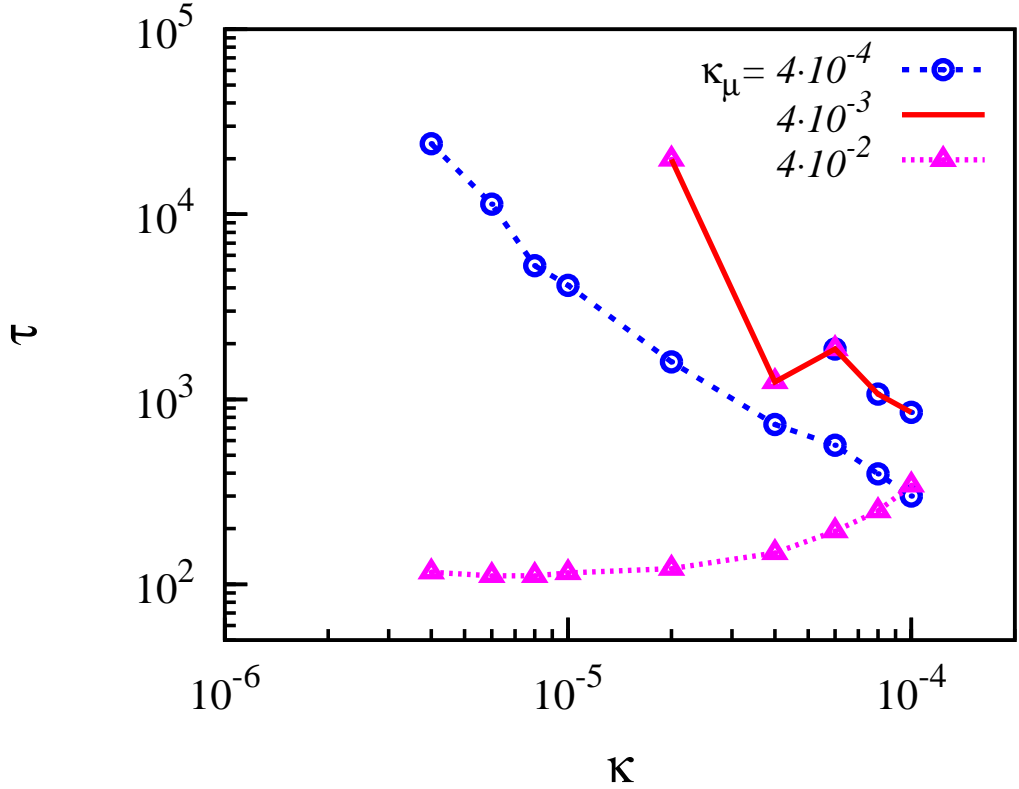}
\caption{The fixation or extinction time, $\tau$, as a function of the
diffusion coefficient $\kappa$ of the Brownian walkers for three values of the
anomalous diffusion coefficient, $\kappa_\mu$, of the L\'evy walkers (with $\mu=1$). Other parameters as
in Figs. \ref{Fig-Phase} and \ref{Fig-Radial}.
The symbols are the same as in Fig.~\ref{Fig-Phase}, referring to which
species survives in the competition; i.e., for $\kappa_\mu = 4 \times 10^{-4}$ ($|\hD_\mu(k_L)|=0.0198$
in Fig. \ref{Fig-Phase}) we depict
the extinction time of Brownian walkers (L\'evy walkers win the competition) and
for $\kappa_\mu = 4 \times 10^{-2}$ ($|\hD_\mu(k_L)|=1.98$) the extinction time of L\'{e}vy walkers
(Brownian walkers win the competition).
For the intermediate value $\kappa_\mu = 4\times 10^{-3}$ ($|\hD_\mu(k_L)|=0.198$) we
observe the two types of outcome, as indicated by the different symbols.
The results are obtained by averaging over 25 realizations.}
\label{Fig-Fix-Time}
\end{figure}

We return to the situation in which both species are initially
present in the system and investigate now the average time
$\tau$ that it takes for one of the two competing species to go
extinct (known as fixation time in the population genetics
jargon). Notice that, since we study a finite system, all
species will become extinct due to large fluctuations at
sufficiently long times. However, in the case of systems
consisting of at least some hundreds of individuals this
happens only after very long time scales. In our simulations we
have never observed such complete extinction and we report only
the fixation time for the species dying out first.
Figure~\ref{Fig-Fix-Time} displays the fixation time, $\tau$,
as a function of the diffusion coefficient, $\kappa$, of the
Brownian walkers for different values of the anomalous
diffusion coefficient, $\kappa_\mu$, of the L\'evy walkers.

For $\kappa_\mu = 4 \times 10^{-4}$, when the L\'evy walkers
always win ($|\hD_\mu(k_L)|=0.0198$, see Fig.~\ref{Fig-Phase}),
the fixation time, $\tau$, decreases when increasing $\kappa$,
i.e., the Brownian walkers die out faster for larger diffusion
coefficient. Namely, larger $\kappa$ means weaker clustering of
the Brownian bugs, thus confirming the observation made in
subsection \ref{sec:radial} of decreased stability for weaker
clustering.

In the largest value $\kappa_\mu = 4 \times 10^{-2}$, which
corresponds to the situation when Brownian walkers win
($|\hD_\mu(k_L)|=1.98$, see Fig.~\ref{Fig-Phase}), the fixation
time, $\tau$, increases for larger values of $\kappa$. The
stronger clustering of Brownian walkers leads to the extinction
of L\'evy ones which form essentially no clusters. The
extinction of the L\'evy walkers is delayed by weaker
clustering of Brownian walkers, i.e., by larger $\kappa$, which
makes the two species more similar.

For the intermediate value $\kappa_\mu = 4 \times 10^{-3}$
($|\hD_\mu(k_L)|=0.198$), when the outcome of the competition
depends on the value of $\kappa$ , the general behavior of
$\tau(\kappa)$ is the same as for $\kappa_\mu = 4 \times
10^{-4}$, i.e., for larger values of $\kappa$ the extinction
time is smaller, increasing rapidly when decreasing $\kappa$.
The local maximum in $\tau(\kappa)$ is associated with the
transition from the regime where L\'evy walkers win due to
their stronger clustering to the regime where the Brownian
walkers win due to their stronger clustering; for this point it
is a random event which of the two species, characterized by a
similar capability of clustering, wins the competition.

As noted at the end of subsection \ref{sec:radial}, for the
intermediate values of $\kappa_\mu$ (e.g., for $\kappa_\mu = 4
\times 10^{-3}$) such that the L\'evy walkers still form
clusters but not too strong ones, and diffusion coefficient
$\kappa$ of the Brownian walkers becomes very small, the
L\'{e}vy walkers do not die out but coexistence of the two
species occurs, i.e., during the accessible simulation time we
do not observe the extinction of neither of the competing
species ($\tau \to \infty$ for $\kappa \to 0$). In this
situation, what happens is that the Brownian walkers form very
strong clusters that the L\'evy walkers are not capable to
invade (because its larger flux out brings them to the death
zones). At the same time, due to the extremely low diffusion
and the high death rate in the inter-clusters space, the
Brownian walkers are not capable to travel away and colonize
the territories that have been occupied by the L\'evy walkers
during the initial cluster formation period due to the
fluctuations.


\subsection{Two-species mean-field description}
\label{sec:theory2}


We now analyze a mean-field deterministic description of the
system, in order to understand the relative stability of
L\'{e}vy {\it versus} Brownian bugs under competition. Denoting
the local density of Brownian walkers by $\rho_B(\bx,t)$ and by
$\rho_L(\bx,t)$ the corresponding to the L\'{e}vy ones, the
dynamics (\ref{1eq}) becomes now:
\BE
\begin{aligned}
\frac{\mathrm{\partial} \rho_B(\bx,t)}{\mathrm{\partial} t} &=& M(\bx,t) \rho_B(\bx,t)  +
D_2 \rho_B(\bx,t) \, ,\\
\frac{\mathrm{\partial} \rho_L(\bx,t)}{\mathrm{\partial} t} &=&
M(\bx,t) \rho_L(\bx,t)  + D_\mu \rho_L(\bx,t) \, , \label{2eqs}
\end{aligned}
\EE
where the difference between birth and death rates is now
$M(\bx,t) \equiv \Delta - G_\bx* \left( \rho_B+\rho_L\right)$.
The notation $D_2$ has been introduced to specify the Brownian
diffusion operator $D_2\equiv\kappa\nabla^2$ and analogously
$D_\mu\equiv\kappa_\mu\nabla^\mu$.

Interaction of the two competing species via Eq. (\ref{2eqs})
is clearly a fully nonlinear problem that can not be addressed
analytically in a complete manner.  But we will see that a
linear analysis around the homogeneous states gives some
insight into the process. We first look for the spatially
homogeneous solutions of Eqs.~(\ref{2eqs}). In this case the
spatial derivatives vanish and there is no difference between
the dynamics of the two species. There exists a family of
steady homogeneous solutions satisfying the condition
$\rho_B+\rho_L=\Delta/\hat G_0$. Thus, we can write the members
of such a family in terms of a parameter $a \in
[-\Delta/(2\hG_0),\Delta/(2\hG_0)]$:
\BE
\rho_B^0 = \frac{\Delta}{2 \hG_0} + a \, , ~~~ \rho_L^0 =
\frac{\Delta}{2 \hG_0} - a \, . \label{homo2}
\EE
The upper boundary of this family ($a =\Delta/(2\hG_0)$)
corresponds to the pure Brownian population, whereas the lower
boundary ($a =-\Delta/(2\hG_0)$) corresponds to the pure
L\'{e}vy population. Intermediate values of $a$ parameterize
different degrees of homogeneous coexistence.

To demonstrate that all this homogeneous family is stable for
sufficiently high values of the diffusivities $\kappa$ and
$\kappa_\mu$ we perturb it with harmonic functions and
calculate the growth rates of such perturbations:
\BE
\begin{aligned}
\rho_B(\bx,t) &=& \rho_B^0 + \delta_B \, e^{\lambda t}
e^{i\bk\cdot\bx} \nonumber \, , \\
\rho_L(\bx,t) &=& \rho_L^0 + \delta_L \, e^{\lambda t}
e^{i\bk\cdot\bx} \, . \label{perturbation}
\end{aligned}
\EE
Linearizing with respect to the small perturbations $\delta_B$
and $\delta_L$, one gets a linear system for which the
solvability conditions give a quadratic equation for $\lambda$,
with two solutions for each value of the set of parameters and
of $k$:
\BE
\lambda_\pm=-\frac{1}{2}A \pm \frac{1}{2}\sqrt{A^2 -4 (B C-D)}
\, , \label{lambda}
\EE
with
\BA
A &=& \Delta \frac{\hG_\bk}{\hG_0} + |\hD_2(k)| + |\hD_\mu(k)| \, , \label{expressions1} \\
B &=& \hG_\bk \left ( \frac{\Delta}{2 \hG_0} + a \right ) + |\hD_2(k)| \, , \label{expressions2} \\
C &=& \hG_\bk \left ( \frac{\Delta}{2 \hG_0} - a \right ) + |\hD_\mu(k)| \, , \label{expressions3} \\
D &=& \hG_\bk^2 \left [ \left ( \frac{\Delta}{2 \hG_0} \right
)^2 - a^2 \right ] \, . \label{expressions4}
\EA
We use the absolute value for $|\hD_2(k)|=-\hD_2(k)=\kappa k^2$
and $|\hD_\mu(k)|=-\hD_\mu(k)=\kappa_\mu k^\mu$ to stress that
these quantities are positive. For sufficiently large
diffusivities $\kappa$ and $\kappa_\mu$, $|\hD_2(k)|$ and
$|\hD_\mu(k)|$ are large and the values of $\lambda_+$ and
$\lambda_-$ are negative (except for the zero mode $\lambda_+
(k = 0) = 0$), meaning that any perturbation applied to a
member of the family of homogeneous solutions decays (except
the neutral perturbations associated to the zero mode, which
transform one of the homogeneous solutions into another one),
and thus any of the homogeneous solutions is stable. No
persistent pattern can appear in the system for large values of
$\kappa$ and $\kappa_\mu$. At each instant the system is in one
of the homogeneous states described by Eqs.~(\ref{homo2}). In
the presence of noise caused by the random birth-death process
(absent from Eqs. (\ref{2eqs})) continuous fluctuations in the
direction of the neutral mode (equivalent to fluctuations in
$a$), will occur transforming one of the homogeneous states
into another one.

When decreasing $\kappa$ or $\kappa_\mu$ ($\hD_2(k)$ or
$\hD_\mu(k)$ approaching zero) the largest of the growth rates
in Eq.~(\ref{lambda}) becomes positive at some values of $k$.
This identifies a pattern-forming instability leading to
periodic modulations of the densities with a characteristic
periodicity given by $2 \pi / k$, similarly to the cases of a
single species described in Sect. \ref{sec:patterns}. Some
information on the character of the instability evolution can
be gained by focussing in its linear growth phase. In
particular one can ask what is the homogeneous combination of
L\'{e}vy and Brownian bugs that becomes unstable first when
decreasing the values of $\kappa$ and $\kappa_\mu$. The
instability condition occurring first is $\lambda_+ > 0$, which
from Eq.~(\ref{lambda}) implies $BC < D$. By using
Eqs.~(\ref{expressions1})-(\ref{expressions4}), this happens if
\BE
|\hD_2(k) \hD_\mu(k)|+\frac{\Delta \hG_\bk}{2
\hG_0}(|\hD_\mu(k)| + |\hD_2(k)|)+a \hG_\bk (|\hD_\mu(k)| -
|\hD_2(k)|) <0 \ . \label{InstabilityCondition}
\EE
When decreasing diffusivities, the instability first occurs for
values of $\bk$ leading to negative values of $\hG_\bk$ and for
$(|\hD_\mu(k)| - |\hD_2(k)|) ~ a > 0$. Due to the linear
dependence in $a$, the earliest instability appears for the
values of $a$ at the extremes of its definition range, i.e.,
for $a = - \Delta / (2 \hat G_0)$ if $|\hD_2(k)|
>|\hD_\mu(k)|$ and for $a = \Delta / (2 \hat G_0)$ if
$|\hD_2(k)| < |\hD_\mu(k)|$. This means that the homogenous
configurations that first become unstable are those populated
by solely L\'{e}vy (when $|\hD_2(k)|>|\hD_\mu(k)|$) or solely
Brownian (when $|\hD_2(k)|<|\hD_\mu(k)|$) walkers. The unstable
mode associated to these instabilities involves only the
L\'{e}vy or the Brownian population, respectively, so that the
pattern that will grow from the unstable state will contain
only that species. Once clusters appear in some part of the
domain, arguments similar to the ones presented in
Sec.~\ref{sec:patterns} indicate that they will dominate the
whole system. The value of $k$ to be used in the above
expressions is the critical one, $k_c$, at which the
instability condition (\ref{InstabilityCondition}) is first
achieved, \ehg{i.e., $k=k_B\approx 4.77901/R$ for the Brownian
homogeneous background (see Table \ref{table1}) should be used
in $|\hat D_2(k)|$, and $k=k_L$, the critical wavenumber $k_c$
for the L\'{e}vy homogeneous background, should be used in
$|\hat D_\mu(k)|$. These critical wavenumbers can be read-off
from Table \ref{table1} at different values of $\mu$.}

Thus, the picture emerging from the mean-field description is
the following: at large diffusion coefficients $\kappa$ and
$\kappa_\mu$ the two types of organisms are essentially the
same and coexistence occurs (until a neutral fluctuation
eliminates irreversibly one of the two species). When
decreasing $\kappa$ and/or $\kappa_\mu$, mixing will be not so
good anymore and different regions of the system may be
occupied by different proportions of bug densities that satisfy
the condition $\rho_B+\rho_L=\Delta/\hat G_0$. By further
decreasing the diffusion coefficients (or increasing $\Delta$
or $R$), some of these regions will encounter an instability.
Which is the first one to occur depends on the sign of
$|\hD_2(k_B)|-|\hD_\mu(k_L)|=\kappa k_B^2 - \kappa_\mu
k_L^\mu$; it is either a pure L\'{e}vy or a pure Brownian
instability, leading to a periodic pattern with periodicity
fixed by $k_B$ or $k_L$, respectively.

As described before, Fig.~\ref{Fig-Phase} shows the outcomes of
the competition process, plotted in a diagram with axes
$|\hD_\mu(k_L)|=\kappa_\mu k_L^\mu$ (with $\mu=1$) and
$|\hD_2(k_B)|=\kappa k^2$. The above arguments predict that the
transition occurs across the main diagonal, which is sketched
in the upper part of the plot. The prediction is not
quantitatively accurate, but it follows qualitatively the trend
of the numerically observed transition, with the proper winning
state above and below it. Although rather crude, the above
linear instability arguments have been adequate to describe
qualitatively the outcome of the competition process. More
importantly, they identify the fluxes or escape rates out of
the clusters, $|\hD_\mu(k_L)|$ and $|\hD_2(k_B)|$, correlated
with cluster width, as the important parameters quantifying the
competitive advantage of one species over the other, in
agreement with the qualitative arguments at the end of
subsection \ref{sec:observations2}.

At this point it is clear that all the above arguments can be
repeated in the case in which we have two types of Brownian
random walkers differing in their diffusion coefficient
$\kappa$, or two species of L\'{e}evy walkers characterized by
different $\mu$ or $\kappa_\mu$. In all these situations the
prediction will be that the species with the smallest value of
the $|\hD(k_c)|$, with $k_c$ the corresponding critical
wavenumber, has a competitive advantage.

\subsection{Limitations of the deterministic description of the two-species system}
\label{sec:limitations2}

The predictions of the previous section for the competition
outcome are only qualitatively correct, as seen is Fig.
\ref{Fig-Phase}, but this is probably more a limitation of the
linear approach used than a failure of the mean-field
description. More important drawbacks arise from the lack of
fluctuations in Eqs. (\ref{2eqs}). As in the single-species
case, this would be relevant to describe the full extinction of
the whole population when $\Delta$ approaches zero. Also, all
the members of the homogeneous family in Eq. (\ref{homo2}) are
predicted to be steady states, but the deterministic treatment
already identifies correctly the change in the parameter $a$ as
a neutral mode, that will fluctuate without restoring force if
noise from the stochastic birth-death process is taken into
account.

\ehg{Competition success in our system is associated to the
pattern formation process and this has been clearly traced back
(Sects. \ref{sec:observations1} and \ref{sec:theory1}) to the
existence of {\sl death zones} arising from the existence of a
finite interaction range $R$. This is well described within the
deterministic approach. However, when interaction range
decreases so that competitive interactions become purely local,
subtle interplays between competition, diffusion and
fluctuations can lead to situations qualitatively different.
For example \citet{Pigolotti2014} show that under local
interactions faster mobility gives competitive advantage, a
mechanism not contained in the corresponding local mean-field
description.}

Perhaps the most relevant deficiency of the deterministic
mean-field description is that it does not capture the
situation when the two species coexist segregated in space. To
describe this situation, a careful consideration of front
propagation and mutual invasion processes is needed. But it can
not be based only in the continuous Eqs. (\ref{2eqs}) because
such study needs to take into account, in addition to the
pinning of the fronts by the periodic structures, the discrete
and fluctuating nature of the system. For example, a continuous
deterministic description of reaction-diffusion processes with
L\'{e}vy diffusion is known to predict an infinite or
accelerating speed of propagating fronts
\citep{delCastillo2003}. But this speed becomes finite in the
corresponding interacting particle model \citep{Brockmann2007}
because of the discrete nature of the particles.

\section{Conclusions and outlook}
\label{sec:conclusions}


In this paper we have summarized results on the dynamics of a
class of stochastic models of competing organisms, in which the
individuals undergo birth and death processes modulated by
competitive interactions with their neighbors, while
simultaneously performing random walks of different types
(Gaussian Brownian motion and L\'{e}vy flights). We have
focused on the way interactions and random motions affect each
other leading to spatial instabilities and extinctions.

Under appropriate conditions, competition produces particle
clustering and the formation of periodic patterns. Periodicity
is of the order of the interaction range $R$, so that it
disappears in purely local interactions ($R\rightarrow 0$).
This process is affected by particle motion in two ways: First
random walks tend to disperse the clusters and homogenize the
system, thus delaying the pattern forming instability. Second,
when different types or species of random walkers are allowed
to compete between them, bugs that form narrower and stronger
clusters get a competitive advantage, in the sense that they
outcompete the others and bring them to extinction.

We have characterized numerically instabilities, patterns and
transitions, and interpreted them with heuristic arguments and
with integrodifferential mean-field models that neglect
fluctuations. In general these continuous models provide a good
description of the individual-based dynamics, in particular the
existence of the pattern forming instabilities and its
wavenumber. We have pointed out some deficiencies, however,
such as incorrect cluster width or coexistence outcomes. These
are mainly associated with the fact that arbitrarily small
densities are allowed in the mean-field description, whereas in
the particle system the minimum number of particles in a region
is either one or zero, with no possible value in between.

The models reviewed here are extremely simplistic, and they can
not be directly compared with realistic biological settings.
Predation, facilitation, parasitism, and many other types of
interaction occur besides competition in real ecosystems, which
would need to be taken into account. Our modeling strategy
should be taken as a mathematical tool that allows a detailed
understanding of basic features of competitive interactions in
combination with random motions. In particular it has clearly
identified the relevance of {\sl death zones} where enhanced
competition occurs, and the importance of the escape rates or
fluxes out of the clusters, quantified by quantities such as
$\hD(k_c)$, in shaping competitive efficiency. \ehg{Despite the
theoretical character of the models we want to mention here
that nonlocal competition processes have been widely discussed
in the context of species competition in niche space
\citep{Barabas2012,Leimar2013} and that some observations of
phenotypic clustering are in general agreement with the pattern
forming instability described here
\citep{Holling1992,Scheffer2006,Segura2012}. Also, plant
competition in semiarid environments provides a framework in
which pattern formation by the competitive instability has been
proposed \citep{Martinez2013}. Here individual plants are not
random walkers, but seed dispersion can be described in a
similar way.}

Even when focusing exclusively in the competition interaction,
some words should be said on the validity of the fundamental
ingredient in the particle model: the basic rates in Eq.
(\ref{r-birthdeath}). In general, when modeling competition for
resources, resource dynamics should be explicitly represented
together with the dynamics of the consumer population
\citep{Schoener1974,Zaldivar2009,Ryabov2011}. Effective
descriptions such as Eq. (\ref{r-birthdeath}) which consider
only the consumers arise in the limit in which resource
dynamics is sufficiently fast
\citep{Schoener1974,Hernandez2009}.

\ehg{Another important ingredient for the development of the
clustering instability is a clear distinction between the
interior and the exterior of the interaction area, so that
there is a well-defined interaction range $R$. This leads to
the existence of `death zones' where particles experience
enhanced competition with neighboring clusters when these are
at a distance in between $R$ and $2R$. We have seen that when
interactions depend on distance via an interaction kernel
$G(\bx)$ the mathematical condition for a sharply defined
interaction range $R$ is the presence of negative components in
the Fourier transform of the interaction kernel, $\hG_\bk$.
When they are present, the death zones, and then the pattern
forming instability, occur (at small mobilities).}

In the context of species competition in niche space early
approaches proposed Gaussian or positively-definite kernels
that precluded this feature
\citep{MacArthur1967,May1972,Roughgarden1979}. But later work
provided more general expressions allowing it
\citep{Schoener1974,Abrams1975,Hernandez2009,Barabas2012,Leimar2013}.
We refer to the last two references for recent reviews of this
species-competition setting. For organisms competing in space
one can envisage several mechanisms leading to a sufficiently
sharp boundary for the competition range, but perhaps the
simplest example arises in plant competition, because of the
finite extent of the roots. The clustering instability has been
reported in that case.

Many open questions remain to be addressed. In particular a
fully nonlinear analytic approach able to describe two-species
competition and coexistence, properly taking into account
particle's discrete nature that greatly affects front
propagation. In the line of including more realistic features,
extensions to other types of organism motion, in particular to
combinations that alternate Brownian and L\'{e}vy displacements
as in \citet{Thiel2012}, are worth to be explored.

\section*{Acknowledgment}
E.H.-G. and C.L. acknowledge financial support from Spanish
MINECO and FEDER through projects INTENSE@COSYP (FIS2012-30634)
and ESCOLA (CTM2012-39025-C02-01). E.H. acknowledges support
from the targeted financing project SF0690030s09, and from the
Estonian Science Foundation grant no. 9462.


\clearpage 

%
%

\bibliographystyle{elsarticle-harv}
\bibliography{Refs-Heinsalu}

%
\clearpage    
\begin{table}[h]
\centering
\begin{tabular}{|c|c|c|c|}
\hline
      $\mu$&   $\nu_c$  & $q_c=k_c R$ & $\delta/R=2\pi/q_c$ \\
\hline
     0.1   &  17.8039  &  5.11619  &   1.22810 \\
     0.5   &  34.1020  &  5.03951  &   1.24679 \\
     1.0   &  76.1997  &  4.94708  &   1.27008 \\
     1.5   &  168.726  &  4.85988  &   1.29287 \\
\hline
     2.0   &  370.384  &  4.77901  &   1.31475 \\
\hline

\end{tabular}
\caption{Critical value $\nu_c$ such that if $\nu>\nu_c$ the
growth rate in Eq. (\ref{growthrate}) becomes positive and
pattern formation occurs, for L\'{e}vy walkers of different
values of $\mu$, and for the Gaussian case which is labeled
with the value $\mu=2$. The selected wavenumber $q_c$ and the
associated pattern periodicity $\delta$ are also displayed. }
\label{table1}
\end{table}

\clearpage

\section*{Figures}

%


%

\end{document}